\documentclass{vgtc}                          

\ifpdf
  \pdfoutput=1\relax                   
  \usepackage{graphicx}                
  \DeclareGraphicsExtensions{.pdf,.png,.jpg,.jpeg} 
\else
  \usepackage{graphicx}                
  \DeclareGraphicsExtensions{.eps}     
\fi%

\graphicspath{{figures/}{pictures/}{images/}{./}} 

\usepackage{microtype}                 
\PassOptionsToPackage{warn}{textcomp}  
\usepackage{textcomp}                  
\usepackage{mathptmx}                  
\usepackage{times}                     
\usepackage{cite}                      
\usepackage{tabu}                      
\usepackage{booktabs}                  

\onlineid{0}

\vgtccategory{Research}

\vgtcinsertpkg

\usepackage{orcidlink}

\title{AR.S.Space: An AR Casual Game for Social Engagement in Work Environments}

\author{Boyuan Chen~\orcidlink{0009-0006-9645-4526}
\and Junkun Long
\and Wenxuan Zheng
\and Yuzheng Wu
\and Ziming Li~\orcidlink{0009-0004-7529-7176}
\and Yue Li~\orcidlink{0000-0003-3728-218X}
\and Hai-Ning Liang~\orcidlink{0000-0003-3600-8955}\thanks{Corresponding author (email: haining.liang@xjtlu.edu.cn)}
}
\affiliation{\scriptsize School of Advanced Technology, Xi'an Jiaotong-Liverpool University, Suzhou, China}

\teaser{
  \centering
  \includegraphics[width=\linewidth]{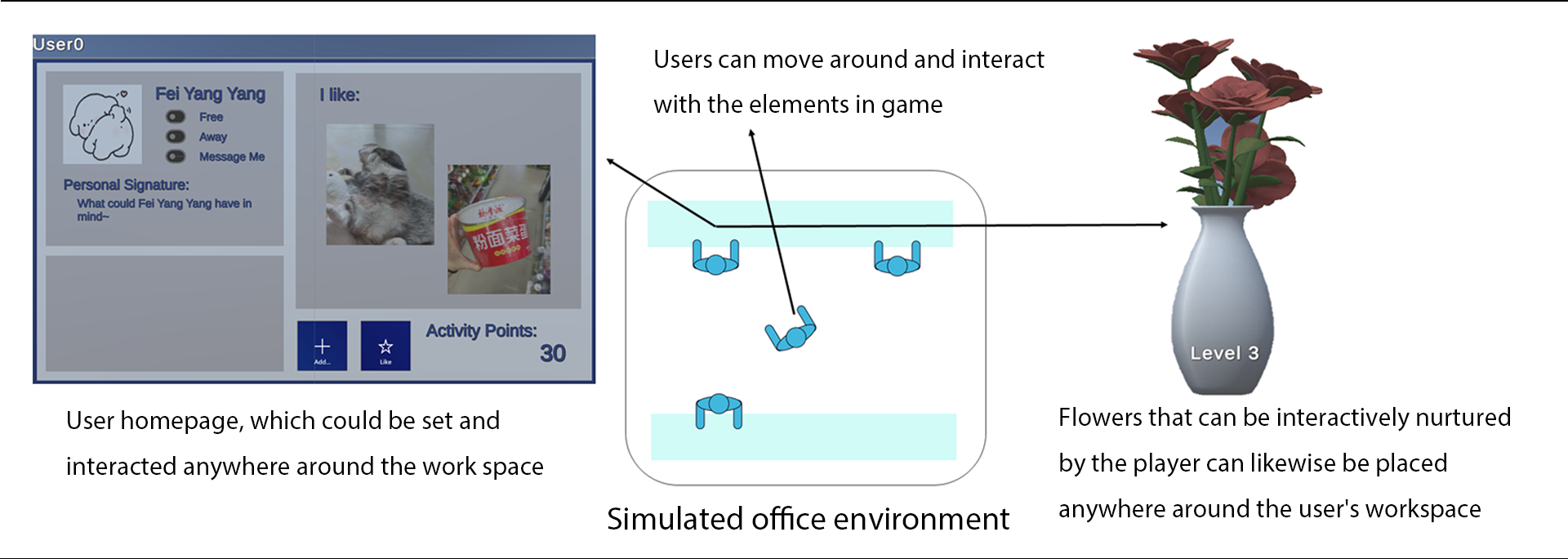}
  \caption{Core features of AR.S.Space}
  \label{fig:Game Interface}
}

\abstract{In social situations, individuals often encounter communication challenges, particularly when adapting to new environments. While some studies have acknowledged the potential of AR social games to aid in effective socialization to some extent, little attention has been given to AR HMD-based games specifically designed to facilitate social interactions. In response, we propose AR.S.Space, an AR HMD-based social game that employs augmented reality features to engage users with virtual social agents through asynchronous communication. The game aims to mitigate the unease associated with initial social interactions and foster long-term connections. To assess its efficacy, a user study was conducted within a specific scenario (an office space), gathering quantitative data and qualitative feedback through questionnaires and interviews. The findings highlight the game's potential to enhance socialization in small-scale environments. Moreover, the study offers valuable design guidelines for future research and the application of AR social games in similar settings.%
} 

\CCScatlist{
 \CCScatTwelve{Human-centered computing}{Mixed / augmented reality}{}{};
 \CCScatTwelve{Human-centered computing}{User Studies}{}{};
}

\begin{document}


\firstsection{Introduction}
\label{sec:Introduction}

\maketitle

Augmented reality (AR) holds the promise of positively influencing various aspects of human life, including the way we think, feel, behave, and communicate\cite{kim2018revisiting}. It has shown tremendous potential in diverse fields, particularly in scenarios where mobility and enriched interactions are crucial. Within the rapidly evolving digital media industry, the integration of AR technology has paved the way for more creative and engaging video games. Among these, social games built on AR head-mounted displays (AR HMD)\cite{kwik2015using} have garnered significant attention from researchers.

Unlike conventional display technologies such as TVs, computers, and smartphones that present a panel in front of the viewer, AR displays offer a see-through capability with an enriched surrounding environment. This unique characteristic provides an appealing new way for individuals to perceive the world and revolutionizes interactions between users, the displayed virtual objects, and the surrounding physical environment\cite{yin2021virtual}. AR HMD offer the advantage of mobility, allowing convenient operation while on the move, while also ensuring the privacy of user interactions. This makes AR HMD an ideal tool for creating social games that can be seamlessly integrated into various real-world scenarios, including office spaces.

In an office setting, socialization plays a critical role in fostering meaningful relationships among colleagues. However, the office environment comes with its unique characteristics that can influence social interactions. People often encounter awkward situations, such as struggling to initiate communication or feeling at a loss for words during socialization\cite{bernstein2018impact}. These challenges can hinder the establishment of strong connections among office employees, affecting collaboration, communication, and overall workplace satisfaction. Thus, it becomes imperative to explore innovative methods that facilitate socialization and create lasting connections within office spaces.

In this paper, we introduce AR.S.Space, an innovative social game based on AR HMD, designed to address the challenges of ice-breaking and establishing long-term connections among employees in an office environment. We draw inspiration from the role of virtual agents in social games to assist players in avoiding awkward social situations and encourage active socialization \cite{li2023faceme}. Our game employs asynchronous messaging to reduce the discomfort of face-to-face interactions, fostering a more enjoyable and effective ice-breaking process. We collect data on player performance and gather feedback on the game's effectiveness through the Game Experience Questionnaire (GEQ) \cite{ijsselsteijn2013game} and a semi-structured interview to ensure its efficacy in enhancing social interactions.

Our research seeks to fill the gap in the existing literature by exploring the potential of AR HMD-based social games, particularly in small-scale social scenarios such as offices or meetings. By leveraging the immersive and interactive capabilities of AR HMD, we aim to provide a more engaging and effective means of ice-breaking and long-term connection-building among colleagues, ultimately contributing to a more vibrant and collaborative office environment. The subsequent sections will detail the design, implementation, and evaluation of AR.S.Space and shed light on its impact on fostering social connections in the office space. Moreover, we propose two research questions to guide our investigation: 
\begin{itemize}
    \item RQ1 - Does the AR HMD social game effectively help players conduct ice-breaking? 
    \item RQ2 - Does the AR HMD social game effectively help players establish long-term relationships? 
\end{itemize}

Through our study, we seek to offer valuable insights and guidelines for future research and application of AR social games tailored to small-scale social environments like offices.

\section{Related Work}
\label{sec: related work}

\subsection{AR Social Games}
Social games represent a distinct category of online games integrated into social networking platforms (SNSs). Leveraging the social features and user data provided by SNSs, these games offer interactive and entertaining experiences for players\cite{hou2011uses,chen2012functional}. The unique characteristics of social games are as follows: First, they are rooted in social platforms, enabling players to engage in communication, cooperation, or competition with others through SNSs, thereby fostering social interaction and impact\cite{hsu2004people,chen2012functional}. Second, players typically utilize their real names and avatars on SNSs, playing games with their real-life network of friends, family, or colleagues, leading to shared identities and a sense of belonging \cite{hou2011uses,chen2012functional}. Last, social games are designed for leisure, simplicity, and ease of access, demanding minimal time and energy investment\cite{paavilainen2013social}. Studies have shown that individuals are more likely to participate in social activities when they share common interests or goals with others\cite{mcmillan1996sense}. Moreover, if people perceive social games as interesting and valuable, they are more inclined to continue using them\cite{wei2014people,shin2008applying}.

The integration of AR technology in social games gained significant traction with the release of Pokemon GO, which appealed to a broad demographic, spanning from children to adults\cite{sonders2016pokemon}. The game's use of location-based AR technology introduced a novel approach to gaming\cite{paavilainen2017pokemon}. Subsequently, researchers have explored various aspects of AR social games, such as constructing location-based AR serious games to educate players about viruses\cite{rapp2018pathomon} and embedding user-generated social media content into physical environments using AR technology\cite{hirsch2022increasing, papangelis2020geomoments}. Furthermore, AR social games have been utilized to promote mental health, aiding individuals with autism\cite{li2023faceme}, and fostering a sense of belonging and peer connection\cite{mittmann2022lina}. While AR social games have been extensively studied and applied, most of these efforts have been focused on specific target groups or commercial applications.

AR holds considerable potential to impact users' social connections\cite{miller2019social}. In AR social games, the researchers proposed AR technology-inspired dimensions of presence, such as content, space, time, and social presence, to give players a uniquely immersive experience \cite{park2022catch}. Studies indicate that utilizing virtual social agents to facilitate socialization can stimulate positive social behavior, especially in individuals with autism spectrum disorder\cite{li2023faceme}, but can reduce the quality of social interactions among people who are in the same place\cite{miller2019social}. However, care must be exercised when implementing social agents in AR social games to avoid unintended negative effects, which can be mitigated by altering the form or state of the social agent.

\subsection{Connection Between Players in a Game}
A key challenge in fostering connections between game players lies in striking a balance between the social and gameplay aspects of the experience\cite{papangelis2017city}. Such connections can be established and maintained through various means, including competition, cooperation, communication, collaboration, conflict resolution, and community building\cite{schell2008art,liang2019fgcs,chen2021cues}. However, the effectiveness of these approaches in enhancing players' perceived enjoyment of the game may vary based on individual preferences and motivations\cite{chen2012functional}. For social games designed to facilitate ice-breaking and establish long-term connections among players, greater emphasis should be placed on elements like cooperation and communication to cultivate harmonious social interactions.

Another critical challenge in fostering connections between players in games involves understanding the cognitive factors that influence players' intentions and behaviors during gameplay. Social cognitive models, such as the Theory of Planned Behavior (TPB) and Social Cognitive Theory (SCT), have been employed to explain the connections between players in games\cite{armitage2000social}. SCT, in particular, has shown success in shaping positive behavioral outcomes in games\cite{hammady2022serious}. By leveraging social cognitive theory, the logic and core mechanisms of player connections in social games can be designed to foster healthy and positive behavioral patterns among players.

However, it is worth noting that most existing studies have primarily focused on social games within large-scale social networks, with limited exploration of social games tailored to small-scale communities, such as office spaces, classrooms, or conferences. Additionally, the effectiveness of AR HMD-based social games has not been extensively investigated in comparison to conventional flat mobile devices. Addressing this research gap, our study aims to design an AR social game specifically tailored to facilitate ice-breaking and build long-term connections in small-scale social scenarios, such as office environments or meetings, potentially surpassing traditional methods like chatting or exchanging business cards.

\section{Social Workspaces and Social Modes}
\label{sec: social scenario and social mode}

\subsection{Office Space}
Sociological theory advocates for the removal of spatial boundaries to promote contact among individuals, fostering collaboration and collective intelligence\cite{festinger1950social}. This argument has been supported by observations in various settings, including university dormitories\cite{festinger1950social}, laboratories\cite{allen1969information}, and co-working spaces\cite{crosina2018start}. As the nature of work evolves, scholars emphasize the removal of spatial boundaries, such as office walls, to create open and borderless offices, aiming to stimulate collaboration and collective intelligence\cite{hight2006collective,parker2016platform,brynjolfsson2014second}. However, studies have revealed that open-architecture workspaces may lead to a natural human response of distancing oneself from office colleagues, resorting to electronic communication instead\cite{bernstein2018impact}. This response is linked to reduced conditions conducive to collaboration, including employee satisfaction\cite{becker1983office}, focus\cite{brookes1972office}, and psychological privacy\cite{sundstrom1982privacy,hundert1969physical}. Consequently, open offices may negatively affect employees' psychological well-being, leading to reduced interactions among office members.

In the context of such office environments with limited face-to-face interactions, newcomers may face challenges in fitting in and establishing connections with their colleagues. The lack of knowledge about coworkers and shared interests makes it difficult to break the ice and build long-term connections. To address these challenges, we propose an AR HMD-based social game that offers a social space with a sense of boundaries. By merging virtual and real scenarios, this game provides a unique opportunity for office workers to reduce the pressure of socialization while fostering connections in a virtual environment, which can protect individuals' psychological privacy and encourage social interactions.

\subsection{Social Mode in Office Spaces}
The socialization process in an office environment can be influenced by Maslow's Hierarchy of Needs Theory, which applies to individuals in office settings\cite{taormina2013maslow}. If employees' social and esteem needs are not met, leading to a lack of a sense of belonging and respect, it can negatively impact their emotions, causing unhappiness, lack of confidence, and uncooperative behaviors. Moreover, Tuckman's team development stage model also provides valuable insights\cite{tuckman1965developmental}. New employees in a team environment generally go through four stages: Formative, Storm, Normative, and Performance stages. In the Formative Stage, employees are curious and nervous, requiring sufficient information to adapt to the new environment. The Storm Stage involves different opinions and potential conflicts, necessitating communication and feedback to resolve issues and adjust attitudes. The Normative Stage sees the development of consensus and trust, requiring engagement and encouragement to enhance a sense of belonging and responsibility. Finally, in the Performance Stage, employees work efficiently and effectively, seeking autonomy, innovation, and development for self-actualization and growth. However, new employees may struggle in the initial stages due to a lack of information and communication feedback, leading to performance deterioration and difficulties in initiating socialization.

Considering the socialization patterns in office spaces, our proposed AR HMD social game focuses on information acquisition and communication for individuals entering a new social environment. By satisfying people's socialization and esteem needs, the game aims to assist employees in breaking the ice and establishing long-term connections in their new workplace. Through information sharing and enhanced communication, the game offers a supportive platform to foster a sense of belonging and social integration among office workers.

\section{AR.S.Space: Design and Implementation}
\label{sec: design and implement AR.S.Space}
We discussed the office environment and the social model in our application scenario. To help people conduct breaking the ice and establish a long-term connection in this scenario, we present an AR HMD social game called AR.S.Space. In the following content, we will introduce the design and implementation of this AR social game.

\subsection{Design Purpose}
\begin{figure*}
    \centering
    \includegraphics[width=\linewidth]{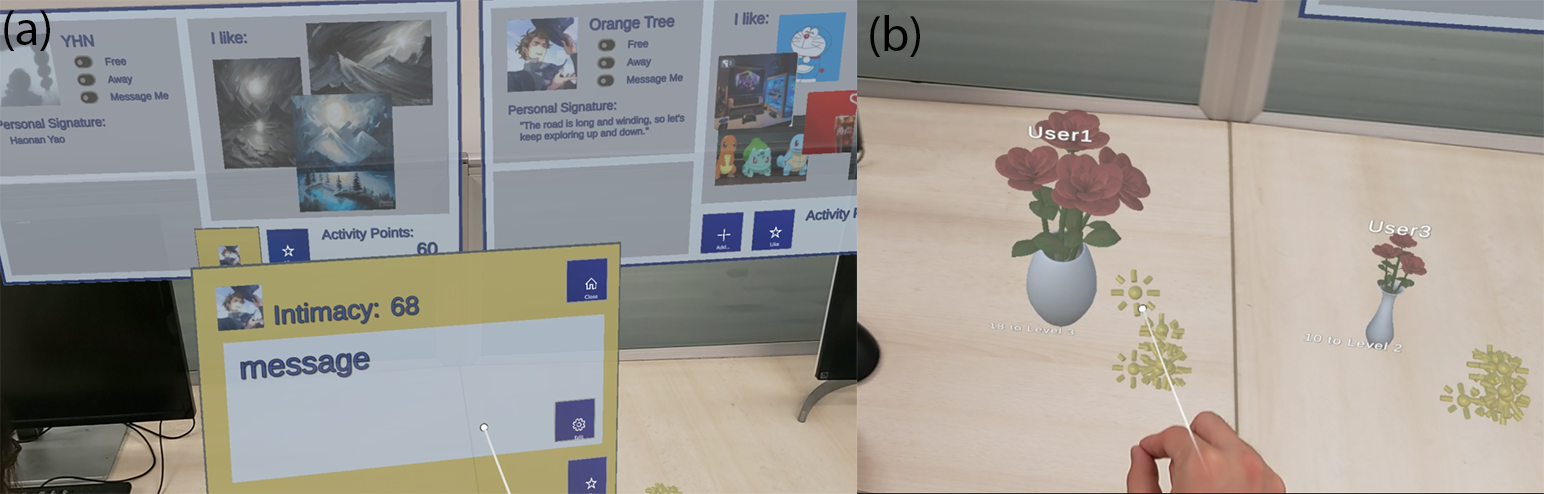}
    \caption{Interact with others on the Homepage and leave messages (a) and users interact with social agents (b)}
    \label{fig: PlayingGame}
\end{figure*}
Currently, within office environments, colleagues primarily establish connections through face-to-face interactions and online chat platforms. However, meeting new colleagues is often hindered by the challenge of identifying shared topics among peers. In addition, social awkwardness is often encountered during face-to-face meetups among young adults, impeding in-person interactions for newcomers. Motivated by this, we introduce AR.S.Space, a game designed to facilitate ice-breaking activities and establish long-term connections. The game's core function is to help individuals quickly get to know each other using virtual environments, reducing the awkwardness of initial meetings and fostering lasting connections. The design purpose of the game focuses on two key aspects.


\textbf{Causal Elements:} The core casual element of this game is a virtual social agent represented by virtual plants. Players need to interact with the virtual plants in order to make the plants grow, which essentially allows for easy gameplay through the design of the nurturing mechanism. Secondly, the asynchronous message and like interactions between players can also increase the intimacy between players, and the design of this kind of relationship nurturing mode between players will also make players feel as little stress as possible when playing the game. Through these casual elements, we want to achieve the design goals of (1) socializing through message boards and asynchronous messaging without being bound or disturbed, as shown in Figure \ref{fig: PlayingGame} (a), and (2) facilitating the development and cultivation of relationships between players through virtual agents, as shown in Figure\ref{fig: PlayingGame} (b)

\textbf{Social Engagement:} The office environment is a relatively closed and quiet environment in which people have difficulty communicating. By setting up personal homepages, we help people get social information about others in a virtual space without having to do so through traditional physical media or face-to-face communication. The purpose of designing personal homepages is to allow people to display the social information they wish to show in the virtual world, thus avoiding the awkwardness of face-to-face communication, and helping players to get to know others quickly.
    
\subsection{System Design}
AR.S.Space is an AR game that supports multiplayer offline asynchronous communication. Each player's actions and changes are recorded on the gaming device so that when the next player picks up the gaming device, they will see the changes made by the previous player. The game is based on HoloLens 2, and all interactions between players are based on that device. The game mechanics of this social game can be summarized as follows. The main gameplay of the game is that players earn affinity and activity points by interacting with other players' personal pages or other players' virtual plants, thus upgrading their virtual plants and increasing their favorability with other players. When players play the game for the first time, they are asked to set up their social personal space. We divided the personal space into two main sections.

\begin{figure*}
    \centering
    \includegraphics[width=\linewidth]{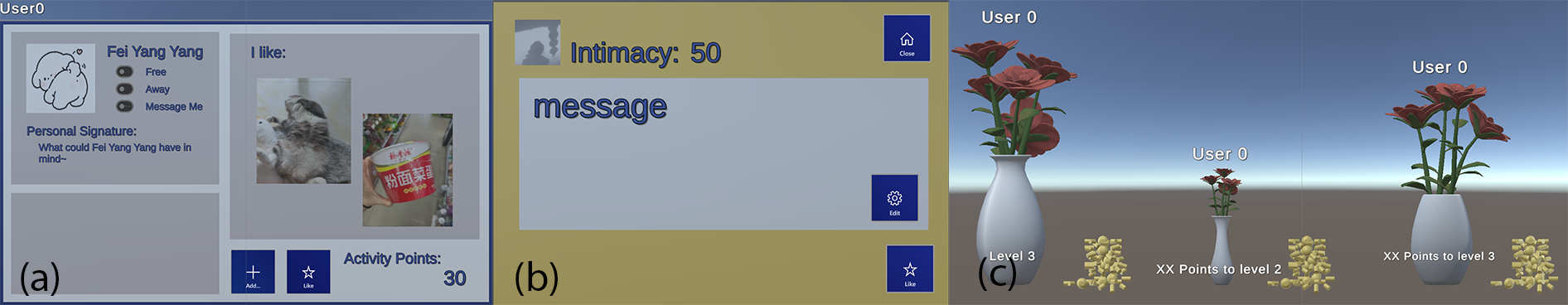}
    \caption{Homepage of users (a), user interface of message (b), the shape of virtual agents at different levels (c)}
    \label{fig: InGame}
\end{figure*}

\textbf{Player's Personal Homepage}, where players can customize the social information they wish to display, as shown in Figure \ref{fig: InGame} (a). Players can customize their avatar, nickname, personal status, personalized signature, and I like. A player's profile can be placed anywhere in the AR space by selecting and dragging it. Other players can like the content of the current player's profile by clicking ``Like" in the profile. When other players like the content of the current player's profile, both the current player and the player who liked the profile will receive 2 activity points. When the activity points increase by 20 points, the virtual plant of players will be upgraded by one level, and the upgrade will make the virtual plant more luxuriant; when the activity points of players increase by 20 points, the virtual plant will be upgraded for a second time, and its appearance will also be changed. The development of virtual plants will increase the stickiness of users and encourage them to play the game more deeply. Players can also leave messages for other players by clicking the Add button on their personal home pages. The game currently only supports text input using the keyboard and does not support voice input because the game's environment is in an office environment, and voice input would disrupt the quietness of the environment to a certain extent. Players can interact with the messages of other players by pressing the ``Like" button, and when the messages are liked, both the current player and the player who left the message will increase the intimacy between the two players by 2 points, as shown in Figure \ref{fig: InGame} (b). This is an effective way to promote positive socialization between the two parties.

\textbf{Virtual Desktop Plants}, which are virtual social agents that we set up for players in the game, as shown in Figure \ref{fig: InGame} (c). Players can likewise place it anywhere in the AR space by selecting and dragging it. Players can interact with the virtual plant asynchronously, thus avoiding as much as possible the awkwardness that can occur when players communicate directly with each other. Sunlight, which players can use to raise plants, is placed next to the virtual plant. Players can place sunlight on the virtual plant by dragging and dropping it, and when another player places sunlight on the current player's plant, both players increase their points by 2 activity points. This way of rewarding both parties encourages people to interact with other people's plants, promoting communication between players. Players are ready to play after setting up their personal homepage and virtual plants.

\subsection{User Interface and Interactions}
AR.S.Space minimizes the learning cost and cognitive load of players through a clear and easy-to-understand interface and easy-to-learn interactions. The block design in the personal interface helps players quickly find areas of interest. On the top left of the personal interface is the player's brief information, showing the player's avatar, nickname, activity status, and personal signature. For players who wish to maintain a certain degree of anonymity, a non-real avatar and name can be used, which fully respects the player's privacy, while the non-real avatar and nickname can also arouse the curiosity of other players. The display of activity status can help other players communicate with the current player at the appropriate time. The player's personalized signature can show the player's personality, and to a certain extent, reduce the recognition cost when other players get to know a person for the first time.

The ``I Like" section on the right side can be customized by each player. It accommodates photos and content related to hobbies, fostering rapid connections by identifying shared interests. When intrigued by another player, clicking ``Add" generates a message on their homepage's bottom left, visible to fellow players. Liking messages or other players' homepages communicates approval. Interaction with other users' content parallels the HoloLens 2 interface, minimizing learning curves and focusing on gameplay. Engaging with virtual plants follows a straightforward pattern: pinching sunlight scattered around plants and placing it on them. This interaction boosts activity points for both participants, fostering social connections by nurturing shared virtual plants.


\subsection{Implementation}
The application was developed in Unity (version 2021.3.8f1c1) for the Universal Windows Platform and deployed on the HoloLens 2 using the Microsoft Mixed Reality Toolkit (version 2.8.0). This combination allowed for intuitive user interactions, such as gesture recognition for manipulating virtual objects and simple clicks and palm gestures for interacting with buttons and menus. The spatial mapping capability enabled users to anchor virtual objects in their real-world environment, enhancing realism and expanding interaction possibilities in the AR space.

\section{User Study}
\label{sec:Evaluation}
To evaluate the effectiveness of our AR social game in facilitating ice-breaking and establishing long-term connections in office environments, we conducted a user study. Quantitative and qualitative data were collected through a simulated office environment with work desks and office supplies. Feedback was obtained using a GEQ questionnaire and semi-structured interviews for assessment.

\subsection{Participants and Procedure}
We randomly recruited four undergraduate or graduate students (two males and two females) as participants on campus to recreate as much as possible what it would be like for people to be new to a new office environment since they did not know each other and they may have been or were about to be experiencing office environments. All participants had experience playing social games, but three of them used AR devices very infrequently, and one of them did not have any experience with AR.
    
Prior to the game, participants completed a pre-experiment questionnaire about their gaming experience and device familiarity. The rules and mechanics of the game were introduced, followed by personalized guidance for setting up their personal space with HoloLens. This phase aimed to familiarize participants with the interactions. They sequentially utilized the AR device, undergoing three rounds of device use to validate ice-breaking and long-term connections. Virtual plant interactions were limited to 2 per player per round, with other actions restricted to 1. A researcher observed and documented interactions. After the game, participants completed the Game Experience Questionnaire (GEQ) to assess their overall gaming experience. Semi-structured interviews gathered more insights on the game's acceptance and reflections on socialization with AR.

\subsection{Results}
In this section, we present the findings derived from the GEQ questionnaire, and in the subsequent subsection, we will elaborate on the user feedback.

\begin{figure}[t]
    \centering
    \includegraphics[width=\linewidth]{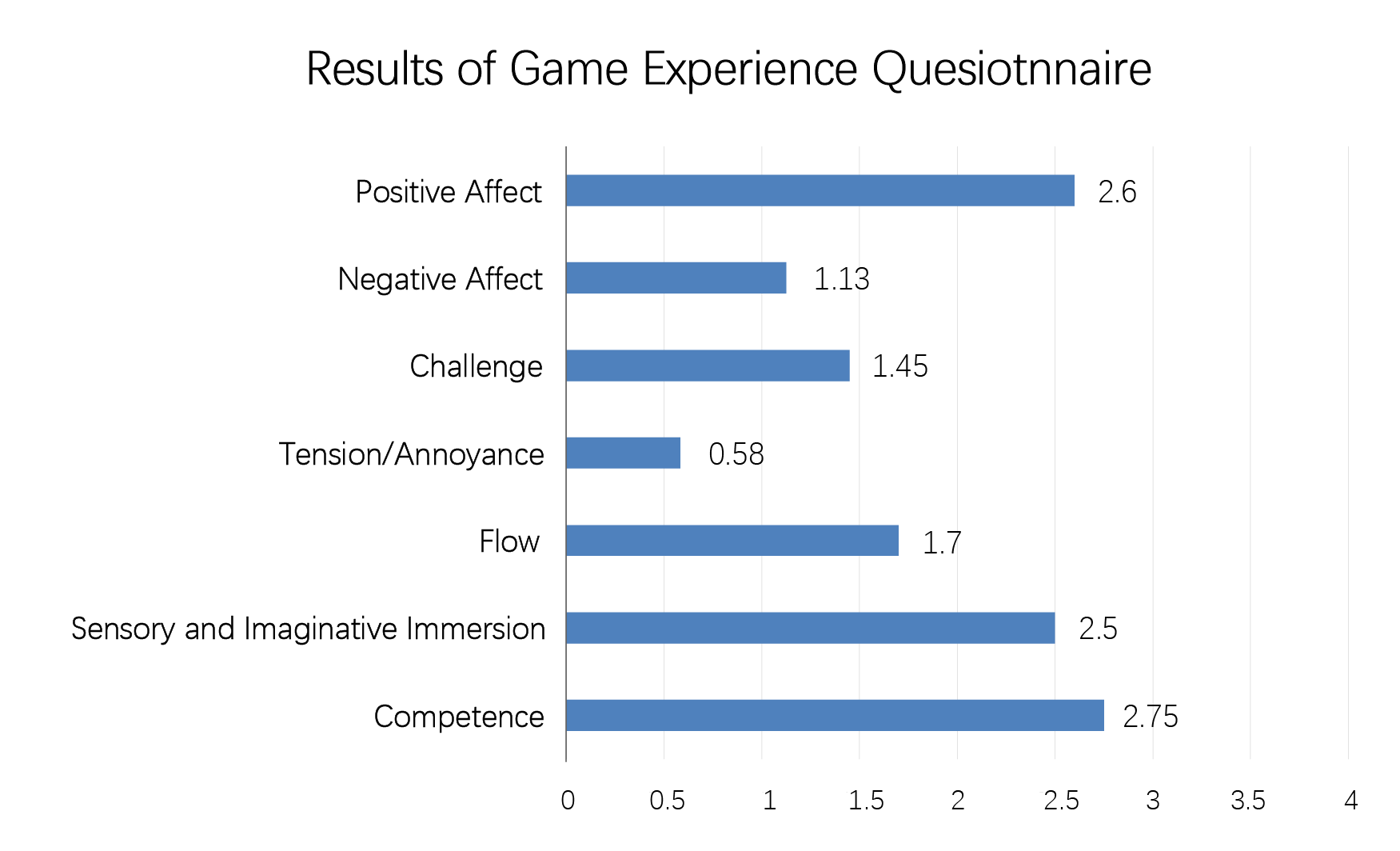}
    \caption{Results of Game Engagement Questionnaire}
    \label{fig: experiment result}
\end{figure}

The results collected through the GEQ questionnaire encompass evaluations across seven distinct dimensions, namely Competence, Sensory and Imaginative Immersion, Flow, Tension/Annoyance, Challenge, Negative affect, and Positive affect. Figure \ref{fig: experiment result} displays the average scores for each evaluation dimension. The outcomes indicate that the game elicited minimal boredom among the players (M=0.58), and they perceived the game to be relatively easy (M=1.45) with low levels of negative affect (M=1.13). These findings suggest that the game is inherently captivating and less demanding to play, aligning well with our initial expectations regarding the game's difficulty level and its entertainment value as a social game. However, it is worth noting that the scores for player competence (M=2.75) and sensory and imaginative immersion (M=2.5) were not as high, indicating areas that warrant improvement to enhance players' sense of competence and their immersive experience, despite the overall positive feedback on the gaming experience.

\subsection{User Feedback}
Feedback from participants during interviews showed unanimous agreement on the effectiveness of displaying personal social information for rapid acquaintance and ice-breaking in small group interactions. Two participants suggested including dynamic social information for stronger long-term connections. Participants appreciated message visibility, promoting interactive engagement.

Three participants praised the virtual plant, effectively reducing face-to-face awkwardness and enhancing the social experience. One participant proposed diverse social proxies, while two recommended more 3D models and gifts for spatial interaction.

Regarding non-game content, participants emphasized the significance of the AR device experience. Two participants expressed concerns about the limited viewing range, especially for AR newcomers. Additionally, two participants suggested improving interaction accuracy and display clarity for distant objects. These insights guide future AR integration improvements in social scenarios.

\section{Discussion}
\label{sec:discussion}
The experimental results demonstrate that our game effectively assists users in ice-breaking and promotes establishing long-term connections between them. This effectively addresses the two research questions that were initially posed. Based on our design discussion and evaluation of AR.S.Space, we present some design recommendations for the potential of game mechanics and scenario-based applications of AR HMD-based social games applied to small-scale communities to help people ice-break and build long-term connections with others in small-scale communities in office-like environments.

\textbf{Conduct Effective Ice-breaking through quick access to other people's information:} One of the main goals of this study is to address the ice-breaking challenge in office environments. Our findings suggest that the inclusion of personalized profiles in the virtual social space of the AR HMD-based game AR.S.Space effectively facilitates the ice-breaking process. The game allows players to quickly learn about each other's interests and preferences through these profiles, thus reducing the awkwardness and uncertainty often accompanying initial social interactions in a new office environment. This feature is a great tool for building connections and starting conversations between people, ultimately promoting a more comfortable and engaging social experience.

\textbf{Relieve social awkwardness through social agents:} Another study finding was that setting up a virtual social agent was more effective in alleviating social awkwardness than face-to-face interactions. Participants indicated that contact with these virtual agents provided a buffer that relieved the stress of direct communication and made the socialization process gradual and less intimidating. The asynchronous nature of interacting with the virtual tabletop plants allowed individuals the flexibility to interact at their own pace, enhancing a sense of control and reducing social discomfort. This feature of the game fits well with Tuckman's model of team development \cite{tuckman1965developmental}, facilitating transitions through the various stages of group formation without overwhelming everyone with immediate face-to-face interactions.

\textbf{Establishing long-term connections requires richer interactive content and dynamic social information of others in AR space:} Research suggests enhancing AR HMD-based social games in offices with richer interactive content and dynamic player interactions. Participants sought diverse 3D AR elements like interactive models and gifts, fostering engagement and sustaining interest. These improvements can deepen connections between players, emphasizing the importance of displaying dynamic social information for mutual understanding. By offering a holistic view of individuals in the virtual environment, the game fosters lasting office connections.

In conclusion, AR.S.Space effectively addresses the research question, providing an ice-breaking and connection-building solution in offices. Personalized profiles expedite information exchange and alleviate initial social discomfort. Social agents ease face-to-face awkwardness, ensuring a seamless, progressive social experience. Enhancing the game's potential for long-term connections involves integrating richer interactive content and dynamic social information in the AR space.

\section{Conclusion}
\label{sec:conclusion}

This study focuses on addressing socialization challenges in office environments by exploring the potential of AR social games. We introduce AR.S.Space, an AR HMD-based social game that creates a virtual social space for office employees to interact and establish connections. Personalized profiles and virtual plants serve as social agents to enhance the social experience and overcome ice-breaking and social awkwardness issues. The integration of personalized profiles in AR.S.Space proved effective in breaking the ice and promoting initial communication. By providing quick access to relevant information about others, uncertainty is reduced, and shared interests are promoted. Additionally, using virtual social agents proved successful in alleviating face-to-face social awkwardness. The asynchronous nature of interacting with these agents allows for gradual engagement, which is especially useful in situations where direct communication can be intimidating. Future research should focus on improving the game further by adding more 3D models and fostering social interaction. Adding voice-based interactions would improve the naturalness of AR interactions. In order to enhance the gaming experience and strengthen player connections, machine learning algorithms can be explored to personalize social agents.


\bibliographystyle{abbrv-doi}

\bibliography{Main}
\end{document}